\documentclass{article}[10pt]

\usepackage{eucal}
\usepackage{natbib}
\usepackage{amsfonts}
\usepackage{mathrsfs}
\usepackage{graphicx}
\usepackage{psfrag}
\usepackage{latexsym}
\usepackage{amsmath}
\usepackage{amstext}
\usepackage{amssymb}
\usepackage{array}
\usepackage{tabularx}
\usepackage{pifont}
\usepackage{psfrag}
\usepackage{color}
\usepackage{manfnt}
\usepackage{wasysym}
\begin{document}
%
\def\bdot{\mbox{\boldmath$\cdot$}}
\def\bcolon{\mbox{\boldmath$\colon$}}
\def\bone{{\bf 1}}
\def\bzero{{\bf 0}}
\def\DIV {\hbox{\rm div}}
\def\Grad{\hbox{\rm Grad}}
\def\sym{\mathop{\rm sym}\nolimits}
\def\dev{\mathop{\rm dev}\nolimits}
\def\Dev{\mathop {\rm DEV}\nolimits}
\def\jmpdelu{{\lbrack\!\lbrack \Delta u\rbrack\!\rbrack}}
\def\jmpudot{{\lbrack\!\lbrack\dot u\rbrack\!\rbrack}}
\def\jmpu{{\lbrack\!\lbrack u\rbrack\!\rbrack}}
\def\jmphi{{\lbrack\!\lbrack\varphi\rbrack\!\rbrack}}
\def\ljmp{{\lbrack\!\lbrack}}
\def\rjmp{{\rbrack\!\rbrack}}
\def\sB{{\mathcal B}}
\def\sU{{\mathcal U}}
\def\sC{{\mathcal C}}
\def\sS{{\mathcal S}}
\def\sV{{\mathcal V}}
\def\sL{{\mathcal L}}
\def\sO{{\mathcal O}}
\def\sH{{\mathcal H}}
\def\sG{{\mathcal G}}
\def\sM{{\mathcal M}}
\def\sN{{\mathcal N}}
\def\sF{{\mathcal F}}
\def\sW{{\mathcal W}}
\def\sT{{\mathcal T}}
\def\sR{{\mathcal R}}
\def\sJ{{\mathcal J}}
\def\sK{{\mathcal K}}
\def\sE{{\mathcal E}}
\def\sS{{\mathcal S}}
\def\sH{{\mathcal H}}
\def\sD{{\mathcal D}}
\def\sG{{\mathcal G}}
\def\sP{{\mathcal P}}
\def\Bnabla{\mbox{\boldmath$\nabla$}}
\def\BGamma{\mbox{\boldmath$\Gamma$}}
\def\BDelta{\mbox{\boldmath$\Delta$}}
\def\BTheta{\mbox{\boldmath$\Theta$}}
\def\BLambda{\mbox{\boldmath$\Lambda$}}
\def\BXi{\mbox{\boldmath$\Xi$}}
\def\BPi{\mbox{\boldmath$\Pi$}}
\def\BSigma{\mbox{\boldmath$\Sigma$}}
\def\BUpsilon{\mbox{\boldmath$\Upsilon$}}
\def\BPhi{\mbox{\boldmath$\Phi$}}
\def\BPsi{\mbox{\boldmath$\Psi$}}
\def\BOmega{\mbox{\boldmath$\Omega$}}
\def\Balpha{\mbox{\boldmath$\alpha$}}
\def\Bbeta{\mbox{\boldmath$\beta$}}
\def\Bgamma{\mbox{\boldmath$\gamma$}}
\def\Bdelta{\mbox{\boldmath$\delta$}}
\def\Bepsilon{\mbox{\boldmath$\epsilon$}}
\def\Bzeta{\mbox{\boldmath$\zeta$}}
\def\Beta{\mbox{\boldmath$\eta$}}
\def\Btheta{\mbox{\boldmath$\theta$}}
\def\Biota{\mbox{\boldmath$\iota$}}
\def\Bkappa{\mbox{\boldmath$\kappa$}}
\def\Blambda{\mbox{\boldmath$\lambda$}}
\def\Bmu{\mbox{\boldmath$\mu$}}
\def\Bnu{\mbox{\boldmath$\nu$}}
\def\Bxi{\mbox{\boldmath$\xi$}}
\def\Bpi{\mbox{\boldmath$\pi$}}
\def\Brho{\mbox{\boldmath$\rho$}}
\def\Bsigma{\mbox{\boldmath$\sigma$}}
\def\Btau{\mbox{\boldmath$\tau$}}
\def\Bupsilon{\mbox{\boldmath$\upsilon$}}
\def\Bphi{\mbox{\boldmath$\phi$}}
\def\Bchi{\mbox{\boldmath$\chi$}}
\def\Bpsi{\mbox{\boldmath$\psi$}}
\def\Bomega{\mbox{\boldmath$\omega$}}
\def\Bvarepsilon{\mbox{\boldmath$\varepsilon$}}
\def\Bvartheta{\mbox{\boldmath$\vartheta$}}
\def\Bvarpi{\mbox{\boldmath$\varpi$}}
\def\Bvarrho{\mbox{\boldmath$\varrho$}}
\def\Bvarsigma{\mbox{\boldmath$\varsigma$}}
\def\Bvarphi{\mbox{\boldmath$\varphi$}}
\def\bA{\mbox{\boldmath$ A$}}
\def\bB{\mbox{\boldmath$ B$}}
\def\bC{\mbox{\boldmath$ C$}}
\def\bD{\mbox{\boldmath$ D$}}
\def\bE{\mbox{\boldmath$ E$}}
\def\bF{\mbox{\boldmath$ F$}}
\def\bG{\mbox{\boldmath$ G$}}
\def\bH{\mbox{\boldmath$ H$}}
\def\bI{\mbox{\boldmath$ I$}}
\def\bJ{\mbox{\boldmath$ J$}}
\def\bK{\mbox{\boldmath$ K$}}
\def\bL{\mbox{\boldmath$ L$}}
\def\bM{\mbox{\boldmath$ M$}}
\def\bN{\mbox{\boldmath$ N$}}
\def\bO{\mbox{\boldmath$ O$}}
\def\bP{\mbox{\boldmath$ P$}}
\def\bQ{\mbox{\boldmath$ Q$}}
\def\bR{\mbox{\boldmath$ R$}}
\def\bS{\mbox{\boldmath$ S$}}
\def\bT{\mbox{\boldmath$ T$}}
\def\bU{\mbox{\boldmath$ U$}}
\def\bV{\mbox{\boldmath$ V$}}
\def\bW{\mbox{\boldmath$ W$}}
\def\bX{\mbox{\boldmath$ X$}}
\def\bY{\mbox{\boldmath$ Y$}}
\def\bZ{\mbox{\boldmath$ Z$}}
\def\ba{\mbox{\boldmath$ a$}}
\def\bb{\mbox{\boldmath$ b$}}
\def\bc{\mbox{\boldmath$ c$}}
\def\bd{\mbox{\boldmath$ d$}}
\def\be{\mbox{\boldmath$ e$}}
\def\bff{\mbox{\boldmath$ f$}}
\def\bg{\mbox{\boldmath$ g$}}
\def\bh{\mbox{\boldmath$ h$}}
\def\bi{\mbox{\boldmath$ i$}}
\def\bj{\mbox{\boldmath$ j$}}
\def\bk{\mbox{\boldmath$ k$}}
\def\bl{\mbox{\boldmath$ l$}}
\def\bm{\mbox{\boldmath$ m$}}
\def\bn{\mbox{\boldmath$ n$}}
\def\bo{\mbox{\boldmath$ o$}}
\def\bp{\mbox{\boldmath$ p$}}
\def\bq{\mbox{\boldmath$ q$}}
\def\br{\mbox{\boldmath$ r$}}
\def\bs{\mbox{\boldmath$ s$}}
\def\bt{\mbox{\boldmath$ t$}}
\def\bu{\mbox{\boldmath$ u$}}
\def\bv{\mbox{\boldmath$ v$}}
\def\bw{\mbox{\boldmath$ w$}}
\def\bx{\mbox{\boldmath$ x$}}
\def\by{\mbox{\boldmath$ y$}}
\def\bz{\mbox{\boldmath$ z$}}
%



\title{The continuum elastic and atomistic viewpoints on the formation volume and strain energy of a point defect}
\author{K. Garikipati\thanks{Assistant professor,
Dept. of Mechanical Engineering, and Program in Applied Physics.
Corresponding author.{\tt krishna@umich.edu}}, M.
Falk\thanks{Assistant professor, Dept. of Materials Science and
Engineering, and Program in Applied Physics.}, M.
Bouville\thanks{Formerly, graduate research assistant, Dept. of
Materials Science and Engineering. Now a post-doctoral fellow at
Institute of Materials Research and Engineering, Singapore.}, B.
Puchala\thanks{Graduate research assistant, Dept. of Materials
Science and Engineering.}, H. Narayanan \thanks{Graduate research
assistant, Dept. of Mechanical Engineering.}\\University of
Michigan, Ann Arbor, Michigan 48109, USA}
\maketitle
%
%










\begin{abstract}
We discuss the roles of continuum linear elasticity and atomistic
calculations in determining the formation volume and the strain
energy of formation of a point defect in a crystal. Our
considerations bear special relevance to defect formation under
stress. The elasticity treatment is based on the Green's function
solution for a center of contraction or expansion in an
anisotropic solid. It makes possible the precise definition of a
formation volume tensor and leads to an extension of Eshelby's
result for the work done by an external stress during the
transformation of a continuum inclusion (\emph{Proc. Roy. Soc.
Lond. Ser. A, \textbf{241}(1226), 376, 1957}). Parameters
necessary for a complete continuum calculation of elastic fields
around a point defect are obtained by comparing with an atomistic
solution in the far field. However, an elasticity result makes it
possible to test the validity of the formation volume that is
obtained via atomistic calculations under various boundary
conditions. It also yields the correction term for formation
volume calculated under these boundary conditions. Using two types
of boundary conditions commonly employed in atomistic
calculations, a comparison is also made of the strain energies of
formation predicted by continuum elasticity and atomistic
calculations. The limitations of the continuum linear elastic
treatment are revealed by comparing with atomistic calculations of
the formation volume and strain energies of small crystals
enclosing point defects.
\end{abstract}

\section{Introduction: Free energy of point defect formation under external stress}
\label{sect1}

When a point defect forms in a crystal there are two main
contributions to the Gibbs free energy of formation,
$\mathscr{G}^\mathrm{f} =
(\mathscr{U}^\mathrm{f}-T\mathscr{S}^\mathrm{f})-\mathscr{W}^\mathrm{ex}$.
The first comes from the internal energy of formation,
$\mathscr{U}^\mathrm{f}$, and the usually negligible entropic term
$-T\mathscr{S}^\mathrm{f}$, where $T$ is the temperature and
$\mathscr{S}^\mathrm{f}$ is the change in entropy. The second
contribution arises due to the work done by the external stress,
and can be formally written as $-\mathscr{W}^\mathrm{ex} =
-\Bsigma^\mathrm{ex}\bcolon\bV^\mathrm{f}$, where
$\Bsigma^\mathrm{ex}$ is the external stress and $\bV^\mathrm{f}$
is the ``formation volume tensor''. The latter quantity is
commonly-used in the materials physics literature
\citep[see][]{Aziz97,Zhaoetal199,Zhaoetal299,dawetal2001} and can
be formally defined as $\bV^\mathrm{f} \equiv
-\partial\mathscr{G}^\mathrm{f}/\partial\Bsigma^\mathrm{ex}$. A
related quantity, the migration volume tensor, $\bV^\mathrm{m}
\equiv
-\partial\mathscr{G}^\mathrm{m}/\partial\Bsigma^\mathrm{ex}$, is
associated with the free energy difference the ground state and
the transition state that the system traverses when the defect
hops between lattice sites. These quantities are of fundamental
interest in materials physics. They influence the kinetics of
transport through the diffusivity, $D = D_0
\exp[-(\mathscr{G}^\mathrm{f} + \mathscr{G}^\mathrm{m})/k_B T]$,
where $k_B$ is the Boltzmann constant.\footnote{Formation plays a
role in self-diffusion of point defects, and in inter-diffusion,
since both phenomena require the formation of a point defect that
subsequently migrates. In doing so it may enable the migration of
a substitutional atom (inter-diffusion).} The thermodynamics of
transport is influenced through the chemical potential, $\mu =
\mu_0 - \Bsigma^\mathrm{ex}\bcolon(\bV^\mathrm{f} +
\bV^\mathrm{m})$, where $\mu_0$ is the potential without stress
effects.

It is natural to attempt a continuum treatment of point defects
since their interaction with stress---a continuum quantity---is of
central interest in this paper. Point defects are modelled as
centers of contraction or expansion within continuum linear
elasticity. While the idea of a formation volume tensor,
introduced above in association with a point defect, is not
standard in continuum mechanics, we place this notion on firm
footing in Section \ref{vform}. Additionally, the introduction of
a formation volume that is energy-conjugate to the stress does
have an analogue in linear elasticity: In a series of seminal
papers on elastic inclusions Eshelby considered ``cutting and
welding'' operations to describe the transformation of an
inclusion within a solid that is itself under external stress
\citep[see, for instance,][]{eshelby57,eshelby61}. He was able to
show that the work done is $\int
\Bsigma^\mathrm{ex}\bcolon\Bvarepsilon^\mathrm{r}\mathrm{d}V$,
where $\Bvarepsilon^\mathrm{r}$ is the transformation strain due
to relaxation, and the integral is over the inclusion. We also
demonstrate in Section \ref{vform} that an extension of Eshelby's
result to point defects can be obtained in a rigorous manner,
leading to the form $-\mathscr{W}^\mathrm{ex} = -
\Bsigma^\mathrm{ex}\bcolon\bV^\mathrm{f}$ introduced above.

The other goal of this communication is to discuss practical
aspects surrounding the evaluation of $\bV^\mathrm{f}$ and
$\mathscr{U}^\mathrm{f}$. On modelling a point defect as a center
of contraction or expansion one can obtain a deformation field
that clearly must bear some relation to $\bV^\mathrm{f}$ and
$\mathscr{U}^\mathrm{f}$. Whatever these relations,
$\bV^\mathrm{f}$ and $\mathscr{U}^\mathrm{f}$ cannot be calculated
unless the strength of the center of contraction or expansion,
represented by a force dipole tensor, is specified in the
elasticity problem. Indeed, in linear elasticity, all elastic
fields scale with the dipole tensor. This quantity must be
obtained from some form of atomistic calculation, as is done in
Section \ref{vform}. However, the linearity of the theory enables
us to show in Sections \ref{vform} and \ref{uform} that even
without quantifying the dipole tensor, rather simple elasticity
calculations reveal the influence of boundary conditions that are
used in the atomistic simulations. A straightforward scaling with
the correct dipole tensor then leads to the actual fields in
elasticity. These calculations are carried out for the isotropic
vacancy in silicon in this preliminary communication. The results
are summarized and placed in the proper context in Section
\ref{sect4}.

\section{The continuum elastic model of a point defect} \label{sect2}

In continuum elasticity a point defect is modelled as an
arrangement of point forces in a center of contraction or
expansion. An associated dipole tensor is defined as\footnote{Both
direct and index notations are used in this paper for brevity and
clarity as deemed necessary.}
\begin{equation}
\bD = \sum\limits_{i=1}^3 \bF_i\otimes 2\bd_i. \label{dipole1}
\end{equation}

\begin{figure}[ht]
\psfrag{A}{\tiny$\bd_1$} \psfrag{B}{\tiny$-\bd_1$}
\psfrag{C}{\tiny$\bd_2$} \psfrag{D}{\tiny$-\bd_2$}
\psfrag{E}{\tiny$\bd_3$} \psfrag{F}{\tiny$-\bd_3$}
\psfrag{G}{\tiny$\bF_1$} \psfrag{H}{\tiny$-\bF_1$}
\psfrag{I}{\tiny$\bF_2$} \psfrag{J}{\tiny$-\bF_2$}
\psfrag{K}{\tiny$\bF_3$} \psfrag{L}{\tiny$-\bF_3$}
\psfrag{M}{\tiny$\be_1$} \psfrag{N}{\tiny$\be_2$}
\psfrag{O}{\tiny$\be_3$} \centering
\includegraphics[width=7cm]{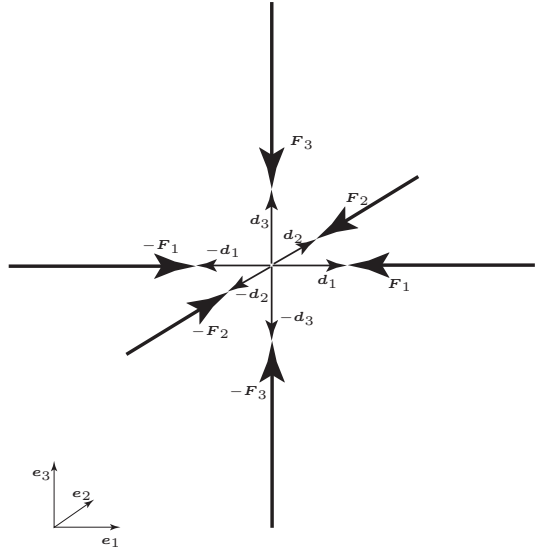}
\caption{An arrangement of point forces leading to an isotropic
dipole---specifically an isotropic vacancy or substitutional
defect that is smaller than the host atoms.} \label{dipolefig}
\end{figure}

\noindent In the dipole illustrated in Figure \ref{dipolefig} the
forces are in equilibrium. Moments are also in equilibrium
provided that $\bD$ is symmetric, which will be assumed here. In
this case $\bD$ can be written with respect to an orthonormal set
of basis vectors $\{\be_1,\be_2,\be_3\}$ that coincide with its
own eigen vectors. In terms of this basis, $\bD$ is therefore
diagonal. Furthermore, if the magnitudes of the point forces are
equal: $\vert\bF_1\vert,\vert\bF_2\vert,\vert\bF_3\vert = F$, and
the magnitudes of the position vectors are also equal:
$\vert\bd_1\vert,\vert\bd_2\vert,\vert\bd_3\vert = d$, then $\bD$
is isotropic and can be written as
\begin{equation}
\bD = D\bone,\;\;\mathrm{where}\; D = \pm 2F d, \label{dipole2}
\end{equation}

\noindent and $\bone$ is the second-order isotropic tensor. The
positive sign in (\ref{dipole2}) holds for centers of contraction
and the negative sign is for centers of expansion as is easily
verified. The isotropy of $\bD$, and $D > 0$ have been assumed for
all numerical calculations in the paper. These choices correspond
to an isotropic vacancy---one in which the magnitudes of
displacements of the nearest-neighbor atoms are equal, or a
substitutional defect with smaller atomic volume than the host
atoms. If $D < 0$, it is an isotropic interstitial or an isotropic
substitutional defect\footnote{Anisotropic interstitials,
substitutional complexes and even vacancies are possible. In
particular, one of the configurations yielded by quantum
mechanical calculations of the relaxation around a vacancy in
silicon is a tetragonal distortion of the four nearest-neighbors.
In the equilibrium position after vacancy formation, all four
atoms are at the same distance from the vacancy. However, the
distances between the four atoms are not the same: The atoms form
two pairs such that the distance separating the atoms in a pair is
smaller than the distance separating each atom in one pair from
the two in the other pair. This is the Jahn-Teller distortion, and
in some instances has been calculated to have a lower formation
energy than the isotropic distortion \citep{puskaetal98,ack98}.}
with larger atomic volume than the host atoms.

Using (\ref{dipole1}), the force distribution due to the dipole is
then written as
\begin{eqnarray*}
\bff(\bx) &=& -2\sum_{i=1}^3 \bF_i\otimes\bd_i
\Bnabla_{x^\prime}\delta(\bx - \bx^\prime)\\
\mathrm{or}\;\;\bff(\bx) &=& -\bD\Bnabla_{x^\prime}\delta(\bx -
\bx^\prime)
\end{eqnarray*}

\noindent where $\delta(\bx-\bx^\prime)$ is the three-dimensional
Dirac-delta function that satisfies $\int
h(\bx)\delta(\bx-\bx^\prime)\mathrm{d}V = h(\bx^\prime)$ for a
sufficiently smooth ($C^\infty$) test function, $h(\bx)$. Since
$\Bnabla_{x^\prime}\delta(\bx - \bx^\prime) =
-\Bnabla_{x}\delta(\bx - \bx^\prime)$, we finally have
\begin{equation}
\bff(\bx) = \bD \Bnabla_{x}\delta(\bx - \bx^\prime).
\label{dipole3}
\end{equation}

The force distribution can be combined with the infinite space
Green's function for anisotropic linear elasticity to obtain the
displacement fields around the defect. The Green's function for
elasticity, $\bG(\bx - \bx^\prime)$, is a second-order tensor. For
an anisotropic, linear medium it satisfies
\begin{equation}
\mathbb{C}_{ijkl}\frac{\partial^2 G_{km}}{\partial x_j \partial
x_l} + \delta_{im}\delta(\bx - \bx^\prime) = 0,
\label{greensfneqn}
\end{equation}

\noindent where $\mathbb{C}_{ijkl}$ is the fourth-order
anisotropic elasticity tensor, $G_{km}$ is the displacement at
$\bx$ along $\be_k$ due to a unit point force acting at
$\bx^\prime$ along $\be_m$, and $\delta_{im}$ is the Kronecker
delta symbol. \citet{barnett72} has applied the Fourier transform
to derive analytic formulae for $G_{ir}$, its first and second
spatial derivatives. These formulae are summarized here:
\begin{equation}
G_{ir} = \frac{1}{4
\pi^2\vert\bx-\bx^\prime\vert}\int\limits_0^\pi
M^\ast_{ir}(\bz(\Psi))\mathrm{d}\Psi, \label{greensfn}
\end{equation}

\begin{equation}
\frac{\partial G_{ir}}{\partial x_s} =
\frac{1}{4\pi^2\vert\bx-\bx^\prime\vert^2}\int\limits_0^\pi\left(-T_s
M^\ast_{ir} + z_s H_{ir}\right) \mathrm{d}\Psi, \label{dgreensfn}
\end{equation}

\begin{eqnarray}
& &\frac{\partial^2 G_{ir}}{\partial x_s\partial x_m} =\nonumber\\
& &\quad
\frac{1}{4\pi^2\vert\bx-\bx^\prime\vert^3}\int\limits_0^\pi\left[2
T_s T_m M^\ast_{ir} - 2\left(z_s T_m + z_m T_s\right)H_{ir} + z_s
z_m A_{ir}\right] \mathrm{d}\Psi. \label{ddgreensfn}
\end{eqnarray}

Letting $\bk$ denote the wave vector in Fourier space, $\bz =
\bk/\vert\bk\vert$ is a unit vector in Fourier space; $\bM^\ast\bM
= \bone$, where $M_{ir}(\bz) = \mathbb{C}_{ijrs}z_j z_s$; $\bT =
(\bx-\bx^\prime)/\vert\bx-\bx^\prime\vert$ is a unit vector in
real space; $\Psi$ is the polar angle in the plane $\bz\cdot\bT =
0$. Using these tensors and vectors $\bH$ and $\bA$ are defined as
\begin{eqnarray*}
H_{ir} &=& \mathbb{C}_{jpnw} M^\ast_{ij} M^\ast_{nr}\left(z_p T_w
+ z_w T_p\right)\\
A_{ir} &=& \mathbb{C}_{jpnw}\left[\left(z_p T_w + z_w
T_p\right)\left(H_{ij} M^\ast_{nr} + M^\ast_{ij}H_{nr}\right) -
2M^\ast_{ij} M^\ast_{nr} T_p T_w\right].
\end{eqnarray*}

\subsection{The displacement and strain fields of a point defect}
\label{sect2.1}

The displacement field of a point defect can be written using
(\ref{dipole3}) and the Green's function:
\begin{displaymath}
\bu^\infty(\bx) = \int\limits_{\mathbb{R}^3}
\bG(\bx-\bx^\prime)\bD\Bnabla_{x^\prime}\delta(\bx^\prime-\bx^{\prime\prime})\mathrm{d}V^\prime,
\end{displaymath}

\noindent where the superscript on the left hand-side signifies
the infinite space solution. Observe that the variable of
integration is $\bx^\prime$. Reverting to indicial notation for
clarity, and using a standard result on derivatives of
distributions, this gives,

\begin{displaymath}
u^\infty_i(\bx) = -\int\limits_{\mathbb{R}^3}
\left[\frac{\partial}{\partial x^\prime_k}
G_{ij}(\bx-\bx^\prime)\right]D_{jk}\delta(\bx^\prime-\bx^{\prime\prime})\mathrm{d}V^\prime.
\end{displaymath}

\noindent Using $\partial G_{ij}(\bx-\bx^\prime)/\partial
x^\prime_k = -\partial G_{ij}(\bx-\bx^\prime)/\partial x_k$, and
the definition of the Dirac-delta function we have

\begin{equation}
u_i^\infty(\bx) = \frac{\partial
G_{ij}(\bx-\bx^{\prime\prime})}{\partial x_k}D_{jk}. \label{disp}
\end{equation}

For an isotropic dipole, $D_{jk} = D\delta_{jk}$, this simplifies
to
\begin{equation}
u_i^\infty(\bx) = D\frac{\partial}{\partial x_j}
G_{ij}(\bx-\bx^{\prime\prime}). \label{dispisotrop}
\end{equation}

The strain field is written as
\begin{equation}
\varepsilon^\infty_{il} = \frac{1}{2}\left(\frac{\partial^2
G_{ij}}{\partial x_k \partial x_l} + \frac{\partial^2
G_{lj}}{\partial x_k \partial x_i}\right)D_{jk}, \label{strain}
\end{equation}

\noindent which, for an isotropic dipole, simplifies to

\begin{equation}
\varepsilon^\infty_{il} = D\frac{1}{2}\left(\frac{\partial^2
G_{ij}}{\partial x_j \partial x_l} + \frac{\partial^2
G_{lj}}{\partial x_j \partial x_i}\right). \label{strainisotrop}
\end{equation}

\noindent Evaluation of (\ref{disp}) and (\ref{strain}) involves
the direct application of (\ref{dgreensfn}) and (\ref{ddgreensfn})
respectively.

\section{The formation volume tensor} \label{vform}

\subsection{The influence of boundaries}
\label{infboun}

The formation volume tensor of a point defect can now be
established using the center of contraction or expansion model.
The results of Section \ref{sect2.1} for infinite crystals are
first extended to finite crystals in order to compare with the
atomistic calculations to be described in Section \ref{sw}. The
development in this subsection was suggested to the authors by
\citet{barnett04} for the \emph{scalar} formation volume,
$\mathrm{tr}[\bV^\mathrm{f}]$. The extension to the full tensor
has not appeared before to the knowledge of the authors.

The formation volume tensor, $V^\mathrm{f}_{{kl}}$, is the sum of
tensorial volume changes due to relaxation of the crystal around
the center of contraction or expansion, $V^\mathrm{r}_{{kl}}$, and
the addition of an atomic volume, $\frac{1}{3}\Omega\delta_{kl}$.
The second term arises since the displaced atom's volume is added
to the crystal. Note the assumption of isotropy associated with
this step. We obtain an expression for $V^\mathrm{f}_{kl}$ below
that, in addition to having other implications, demonstrates that
$V^\mathrm{r}_{{kl}}$, and therefore $V^\mathrm{f}_{{kl}}$, are
well-defined quantities.

For a defect in a finite crystal, $B_\mathrm{crys}$, we define
\begin{equation}
V^\mathrm{f}_{{kl}} \equiv
\underbrace{\int\limits_{B_\mathrm{crys}}\varepsilon_{kl}\mathrm{d}V}_{V^\mathrm{r}_{{kl}}}
\;+\; \frac{1}{3}\Omega\delta_{kl}, \label{barnett1}
\end{equation}

\noindent where $\varepsilon_{kl}$ is the strain field in the
finite crystal due to the center of contraction or expansion with
the appropriate value of the dipole. Using the stress-strain
relation, this is
\begin{equation}
V^\mathrm{f}_{{kl}} =
\mathbb{S}_{klij}\int\limits_{B_\mathrm{crys}}\sigma_{ij}\mathrm{d}V
+ \frac{1}{3}\Omega\delta_{kl} \label{barnett2}
\end{equation}

\noindent where $\mathbb{S}_{klij}$ is the constant compliance
tensor satisfying $\mathbb{S}_{klij}\mathbb{C}_{ijmn} =
\mathbb{I}_{klmn}$, the fourth-order symmetric identity tensor.
Observe that
\begin{equation}
\int\limits_{B_\mathrm{crys}}\sigma_{ij} \mathrm{d}V =
\int\limits_{B_\mathrm{crys}}\left[\frac{\partial\left(x_i\sigma_{mj}\right)}{\partial
x_m} - x_i\frac{\partial\sigma_{mj}}{\partial x_m}
\right]\mathrm{d}V. \label{barnett3}
\end{equation}

\noindent Using (\ref{dipole3}) the equilibrium equation for the
center of contraction or expansion is written as

\begin{equation}
\frac{\partial\sigma_{jm}}{\partial x_m} +
D_{jm}\frac{\partial\delta(\bx-\bx^\prime)}{\partial x_m} = 0.
\label{barnett4}
\end{equation}

\noindent Combining (\ref{barnett3}) and (\ref{barnett4}) and
using the symmetry of $\Bsigma$ we have,

\begin{displaymath}
\int\limits_{B_\mathrm{crys}}\sigma_{ij} \mathrm{d}V =
\int\limits_{B_\mathrm{crys}}\left[\frac{\partial\left(x_i\sigma_{mj}\right)}{\partial
x_m} + x_i D_{jm}\frac{\partial\delta(\bx-\bx^\prime)}{\partial
x_m}\right] \mathrm{d}V.
\end{displaymath}

\noindent Using the standard result for spatial derivative of the
Dirac-delta function, the symmetry of $\Bsigma$, and the
Divergence Theorem,
\begin{displaymath}
\int\limits_{B_\mathrm{crys}}\sigma_{ij} \mathrm{d}V =
\int\limits_{S_\mathrm{crys}}x_i\sigma_{jm} n_m \mathrm{d}A -
D_{jm}\delta_{im},
\end{displaymath}

\noindent where $S_\mathrm{crys}$ is the surface of the crystal.
Substituting this equation in (\ref{barnett2}) we have
\begin{equation}
V^\mathrm{f}_{{kl}} = \underbrace{-\mathbb{S}_{klij}D_{ji}
+\mathbb{S}_{klij}\int\limits_{S_\mathrm{crys}}x_i
\sigma_{jm}n_m\mathrm{d}A}_{V^\mathrm{r}_{{kl}}}\; +\;
\frac{1}{3}\Omega\delta_{kl} . \label{barnett5}
\end{equation}

This result has three important, related, consequences:
(\romannumeral 1) The relaxation and formation volume tensors for
a crystal with traction-free boundaries depend only on the
strength of the center of contraction or expansion, represented
here by the dipole $D_{ji}$, elasticity of the crystal, and the
atomic volume (for the case of formation volume). In particular,
neither the shape and size of the crystal nor the location of the
center of contraction or expansion inside the crystal influence
the outcome. (\romannumeral 2) The difference in
$V^\mathrm{r}_{{kl}}$ between the case with traction-free boundary
and one with arbitrary boundary conditions can be evaluated. For
pure traction boundary conditions this is trivial. With
displacement boundary conditions the elasticity problem must be
solved first, possibly by numerical methods. (\romannumeral 3)
Formation and relaxation volume tensors, unorthodox notions in
continuum mechanics, can be defined in a precise fashion.

\subsection{A thermodynamic basis for $\bV^\mathrm{f}$}
\label{disclatt}

The center of contraction or expansion model for a point defect
thus leads to a relation for $\bV^\mathrm{f}$ that can be
evaluated if the stress field of the crystal is known. However,
the thermodynamic basis of this quantity has not been clarified
beyond the formal relation $\bV^\mathrm{f} =
-\partial\mathscr{G}^\mathrm{f}/\partial \Bsigma^\mathrm{ex}$. In
this subsection it is demonstrated that a result of
\citet{eshelby57}, established for continuum inclusions, also
holds in the center of contraction or expansion limit and provides
such a thermodynamic basis.

We begin with the continuum result for the work of interaction
between the traction, $\Bsigma^\mathrm{ex}\bn$, applied at the
boundary of a crystal, and the deformation induced by a
transforming inclusion in the crystal. As alluded to in Section
\ref{sect1}, Eshelby showed that this work of interaction has the
form
\begin{equation}
\mathscr{W}^\mathrm{ex} =
\int\limits_{B_\mathrm{inc}}\Bsigma^\mathrm{ex}\bcolon\Bvarepsilon^\mathrm{r}\mathrm{d}V
\label{eshelby1}
\end{equation}

\noindent where $B_\mathrm{inc}$ is the region occupied by the
inclusion before it undergoes the \emph{stress-free} relaxation
strain $\Bvarepsilon^\mathrm{r}$ \citep{eshelby57}. This is the
strain in the inclusion if its boundary, $S_\mathrm{inc}$, is
traction-free. If the inclusion is small enough that
$\Bsigma^\mathrm{ex}$ is constant over $B_\mathrm{inc}$, then
following (\ref{barnett1}) and defining
$\bV^\mathrm{r}_\mathrm{inc} \equiv \int_{B_\mathrm{inc}}
\Bvarepsilon^\mathrm{r}\mathrm{d}V$ allows us to rewrite
(\ref{eshelby1}) as
\begin{equation}
\mathscr{W}^\mathrm{ex} =
\Bsigma^\mathrm{ex}\bcolon\bV^\mathrm{r}_\mathrm{inc}.
\label{eshelby2}
\end{equation}

\begin{figure}[ht]
\centering
\includegraphics[width=5cm]{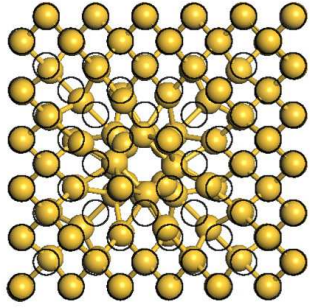}
\vspace{-2cm}
\caption{A vacancy in silicon viewed along the
$\langle 1\,0\,0\rangle$ direction. The open circles mark
positions of atoms in the perfect crystal; i.e., before vacancy
formation. Displacements have been scaled $10\times$ for clarity.}
\label{vacancyfig}
\end{figure}

A seemingly obvious step is to link $\bV^\mathrm{r}_\mathrm{inc}$
and $\bV^\mathrm{r}$ for the point defect. Then Eshelby's result
would apply to the work of relaxation around a point defect, and
the work of formation would be
$\Bsigma^\mathrm{ex}\bcolon(\bV^\mathrm{r} +
\frac{1}{3}\Omega\bone)$. Introducing this work term into the
Gibbs free energy would lead to $\bV^\mathrm{f} =
-\partial\mathscr{G}^\mathrm{f}/\partial \Bsigma^\mathrm{ex}$.
Such an approach has been adopted by \citet{Azizetal91}. In our
notation their expression is $\mathscr{W}^\mathrm{ex} =
\Bsigma^\mathrm{ex}\bcolon\Bvarepsilon^\mathrm{T}\cdot
\mathrm{Vol}(B_\mathrm{inc})$, where $\Bvarepsilon^\mathrm{T}$ is
the uniform transformation strain undergone by a ``sub-system'',
$B_\mathrm{inc}$. However, there are technical difficulties in
making such a direct association: (\romannumeral 1) The boundary
between the continuum inclusion and the solid in which it is
embedded is well-defined, but it is not possible to precisely
demarcate a lattice point defect in this manner. For example,
Figure \ref{vacancyfig} shows a vacancy in a silicon crystal where
the $\langle 1\,0\,0\rangle$ direction is perpendicular to the
plane of the figure (this is a visualization of one of our
atomistic calculations). It is evident that no precise interface
exists between the defect and the perfect lattice. Indeed, in the
above expression used by Aziz and co-workers the extent of the
``sub-system'' is not made precise beyond a statement that it
contains the atoms ``involved in the fluctuation'' of the
transformation. The transformation strain is also not made
precise. (\romannumeral 2) The arguments that Eshelby used in
arriving at (\ref{eshelby1}) require that the continuum assumption
holds even in the neighborhood of the inclusion. The discreteness
of the lattice clearly negates this assumption in the neighborhood
of a point defect.

The center of contraction or expansion model of the point defect
in continuum linear elasticity enables a circumvention of these
difficulties: Consider a finite crystal with traction-free
boundaries that is also stress-free in its interior. As before,
$B_\mathrm{crys}$ is the domain of the crystal and its boundary is
$S_\mathrm{crys}$. Let an external traction,
$\Bsigma^\mathrm{ex}\bn$ be applied at $S_\mathrm{crys}$. Also let
$\Bvarepsilon^\mathrm{ex}$ and $\bu^\mathrm{ex}$ be the related
strain and displacement fields in $B_\mathrm{crys}$. Now let a
point defect form in the interior of $B_\mathrm{crys}$ with
corresponding dipole tensor $\bD$. Let the stress, strain and
displacement fields arising from this defect be
$\Bsigma^\mathrm{F}$, $\Bvarepsilon^\mathrm{F}$ and
$\bu^\mathrm{F}$ in the case that $S_\mathrm{crys}$ is
traction-free. Also recall that according to the linear theory of
elasticity the fields actually obtained now are
$\Bsigma^\mathrm{ex} + \Bsigma^\mathrm{F}$,
$\Bvarepsilon^\mathrm{ex} + \Bvarepsilon^\mathrm{F}$ and
$\bu^\mathrm{ex} + \bu^\mathrm{F}$. The quantity of interest is
the work done by the external stress on the crystal in the process
of defect formation:
\begin{displaymath}
\mathscr{W}^\mathrm{ex} =
\int\limits_{S_\mathrm{crys}}u^\mathrm{F}_i(\sigma^\mathrm{ex}_{ij}
n_j)\mathrm{d}A.
\end{displaymath}

\noindent Using the Divergence Theorem and integration by parts
this is
\begin{displaymath}
\mathscr{W}^\mathrm{ex}=
\int\limits_{B_\mathrm{crys}}\left(u^\mathrm{F}_i\sigma^\mathrm{ex}_{ij,j}
+
\varepsilon^\mathrm{F}_{ij}\sigma^\mathrm{ex}_{ij}\right)\mathrm{d}V,
\end{displaymath}

\noindent where the symmetry of $\Bsigma^\mathrm{ex}$ has been
used to restrict the displacement gradient to its symmetric part,
the strain. Since $\Bsigma^\mathrm{ex}$ is divergence-free in
$B_\mathrm{crys}$ the first term in the above integrand vanishes.
The second term is rewritten using the stress-strain relations to
give
\begin{displaymath}
\mathscr{W}^\mathrm{ex}= \int\limits_{B_\mathrm{crys}}
\sigma^\mathrm{F}_{ij}\varepsilon^\mathrm{ex}_{ij}\mathrm{d}V.
\end{displaymath}

\noindent Invoking the symmetry of $\Bsigma^\mathrm{F}$ and using
integration by parts,
\begin{displaymath}
\mathscr{W}^\mathrm{ex}= \int\limits_{S_\mathrm{crys}}
u^\mathrm{ex}_i \sigma^\mathrm{F}_{ij} n_j\mathrm{d}A -
\int\limits_{B_\mathrm{crys}}
u^\mathrm{ex}_i\sigma^\mathrm{F}_{ij,j} \mathrm{d}V.
\end{displaymath}

\noindent Since $\Bsigma^\mathrm{F}$ is obtained as the solution
to the traction-free boundary case, the first term in this
integrand vanishes. The second term is rewritten using
$\sigma^\mathrm{F}_{ij,j} + D_{ij}\delta_{,j}(\bx-\bx^\prime) =
0$, for a defect at $\bx = \bx^\prime$, to give
\begin{displaymath}
\mathscr{W}^\mathrm{ex}= \int\limits_{B_\mathrm{crys}}
u^\mathrm{ex}_i D_{ij}\frac{\partial}{\partial x_j}\delta(\bx -
\bx^\prime) \mathrm{d}V.
\end{displaymath}

\noindent Using the standard result for derivative of the
Dirac-delta function this gives
\begin{displaymath}
\mathscr{W}^\mathrm{ex}=
 -u^\mathrm{ex}_{i,j}(\bx^\prime)
D_{ij}.
\end{displaymath}

\noindent For a symmetric dipole tensor (cf Section \ref{sect2})
\begin{displaymath}
\mathscr{W}^\mathrm{ex}=
 -\varepsilon^\mathrm{ex}_{ij}(\bx^\prime)
D_{ij}.
\end{displaymath}

\noindent From the stress-strain relations in the form
$\varepsilon^\mathrm{ex}_{ij} =
\mathbb{S}_{ijkl}\sigma^\mathrm{ex}_{kl}$ and the major and minor
symmetries of $\mathbb{S}_{ijkl}$ we have
\begin{displaymath}
\mathscr{W}^\mathrm{ex}=
 \sigma^\mathrm{ex}_{kl}(\bx^\prime)\left(-\mathbb{S}_{klij}
D_{ji}\right).
\end{displaymath}

\noindent Finally, recalling (\ref{barnett5}) and noting that
$\sigma_{jm}n_m$ in the boundary integral of that equation is now
replaced by $\sigma^\mathrm{F}_{jm}n_m = 0$ we have
\begin{equation}
\mathscr{W}^\mathrm{ex}= \sigma^\mathrm{ex}_{kl}
V^\mathrm{r}_{{kl}}.\label{garikipati1}
\end{equation}

On including the contribution of the displaced atom as in Section
\ref{infboun}, this result can be extended to the work done in
defect formation:
\begin{equation}
\mathscr{W}^\mathrm{ex}= \sigma^\mathrm{ex}_{kl}\left(
V^\mathrm{r}_{{kl}} + \frac{1}{3}\Omega\delta_{kl}\right) =
\sigma^\mathrm{ex}_{kl} V^\mathrm{f}_{kl} .\label{garikipati2}
\end{equation}

The use of (\ref{garikipati2}) for $\mathscr{W}^\mathrm{ex}$ in
$\mathscr{G}^\mathrm{f}$ lends rigor to the thermodynamic
definition $\bV^\mathrm{f} \equiv
-\partial\mathscr{G}^\mathrm{f}/\partial\Bsigma^\mathrm{ex}$
(Section \ref{sect1}). By extension through $\bV^\mathrm{r} =
\bV^\mathrm{f} - \frac{1}{3}\Omega\bone$ this development also
provides a thermodynamic basis for $\bV^\mathrm{r}$. Thus the
center of contraction or expansion model makes the notion of
formation and relaxation volume tensors precise, and also provides
a thermodynamic basis for these quantities.

\noindent\textbf{Remark 1}: \citet{NowickHeller63} have defined a
volume tensor analogous to $\bV^\mathrm{r}$. Their expression is
$v\lambda_{ij}$, where $v$ is the atomic volume and $\lambda_{ij}$
is what they call an ``elastic dipole tensor''. The volume tensor
$v\lambda_{ij}$ is conjugate to the external stress via the
interaction energy. However, $\lambda_{ij}$ is ``equal to the
average strain per mole fraction of defects all aligned in a
particular orientation''. This definition is therefore
considerably different from our treatment. To our knowledge
neither this definition of a volume tensor nor that of
\citet{Azizetal91} has been rigorously \emph{derived} in a
kinematic or thermodynamic sense. Our search of the literature on
point defects has not uncovered either a kinematic definition of
$\bV^\mathrm{r}$ as in Section \ref{infboun} or a thermodynamic
basis for it as in Section \ref{disclatt}.

\subsection{Atomistic calculations for the formation volume and dipole}
\label{sw}

The solution for any elastic field around the center of
contraction or expansion depends on the dipole strength, $\bD$,
which must be known to fully specify the elasticity problem. Thus,
the formation volume, $\bV^\mathrm{f}$, cannot be obtained from
elasticity alone. The dipole, $\bD$, and hence $\bV^\mathrm{f}$,
must be obtained from atomistic calculations using empirical
interatomic potentials, tight binding or \emph{ab initio} methods.
A comparison between such atomistic calculations and a simple
continuum result allows a determination of $\bD$ in closed-form as
shown in Section \ref{vrtensor}. With the value of $\bD$ thus
known, continuum elasticity can be used for a variety of
calculations involving the elastic fields around the point defect.
In Section \ref{boundefsymm} we show that the elasticity results
of Section \ref{infboun} also provide important information
regarding the validity of atomistic calculations for
$\mathrm{tr}[\bV^\mathrm{f}]$. The results of Section
\ref{sect3.2.2} similarly demonstrate the importance of
interactions between the vacancy, elastic fields and boundary
loading mechanism in determining the strain energy of formation,
$\mathscr{U}^\mathrm{f}$.

We have undertaken atomistic calculations of vacancy formation
using the Stillinger-Weber interatomic potential \citep{sw85}.
Although this empirical potential is a rough approximation of
silicon, particularly at the defect, it is suitable for the
purpose of checking consistency between continuum elastic and
atomistic predictions for $\bV^\mathrm{f}$ and
$\mathscr{U}^\mathrm{f}$ in the far-field.

The Stillinger-Weber potential consists of two- and three-body
terms governing the interactions of N atoms:
\begin{eqnarray}
    \Phi (1,...,N) &=& \sum_{i<j}^N v_{2}(r_{ij}) + \sum_{i<j<k}^N v_{3}(\br_i,\br_j,\br_k) \label{sw1}\\
    v_{2}(r_{ij}) &=&  e f_{2}\left(\frac{r_{ij}}{\xi}\right) \label{sw2}\\
    v_{3}(\br_i,\br_j,\br_k) &=& e f_{3}
    \left(\frac{\br_i}{\xi},\frac{\br_j}{\xi},\frac{\br_k}{\xi}\right)\label{sw3}
\end{eqnarray}

\noindent where energy and length units $e$ and $\xi$ are chosen
to give $f_{2}$ a minimum value of $-1$ if its argument is
$2^{1/6}$. The two-body function, $f_{2}$, depends only on the
distance $r_{ij} = \vert\br_i - \br_j\vert$ between a pair of
atoms with position vectors $\br_i$ and $\br_j$ and has a cut-off
at $r=a$ without discontinuities in any derivatives with respect
to $r$:
\begin{equation}
f_{2}(r) =
  \left\{\begin{array}{c@{\quad:\quad}l}
    A(Br^{-p}-r^{-q}) \exp(\frac{1}{r-a})& \mbox{if}\;r<a, \\
    0  & \mbox{if}\;r\geq a
  \end{array}\right.
\label{sw4}
\end{equation}

\noindent The three-body function, $f_{3}$, depends on the scalar
distances between the three atoms and also on the angle subtended
at the vertices of the triangle formed by the three atoms. In the
following relations $\theta_{jik}$ is the angle subtended at
vertex $i$ between atoms with position vectors $\br_j$ and
$\br_k$:
\begin{eqnarray}
    f_{3} \left(\frac{\br_i}{\xi},\frac{\br_j}{\xi},\frac{\br_k}{\xi}\right)&=& h\left(\frac{r_{ij}}{\xi},\frac{r_{ik}}{\xi},\theta _{jik}\right)+h\left(\frac{r_{ji}}{\xi},\frac{r_{jk}}{\xi},\theta _{ijk}\right)\nonumber\\
    &&\hspace{3cm}+\;h\left(\frac{r_{ki}}{\xi},\frac{r_{kj}}{\xi},\theta _{ikj}\right) \\
    h(r_1,r_2,\theta) &=& \left\{
    \begin{array}{c@{\;:\;}l}
    \lambda \exp(\frac{\gamma}{r_1-a} + \frac{\gamma}{r_2-a}) \left(\cos\theta+\frac{1}{3}\right)^{2}& r_1\;\mbox{and}\; r_2<a \\
    0 &r_1\;\mbox{or}\; r_2 \geq a. \\
    \end{array}\right.
    \label{sw5}
\end{eqnarray}

\noindent The equilibrium bond angle in the tetrahedral structure
of silicon satisfies $\cos\theta = -1/3$. The energetic
contribution of any bond angle distortions are thus represented by
the trigonometric term in (\ref{sw5}). The lattice spacing and
bond energy of silicon at $0$ K are obtained with $\xi = 0.20951$
nm and $e = 2.346\;\mbox{eV}$.

The optimized set of parameters  for this function is:
\begin{alignat*}{2}
    & A=7.049556277,\quad &&B=0.6022245584 \\
    & p=4,\ q=0, &&a=1.80 \\
    & \lambda=21.0, && \gamma=1.20.
\end{alignat*}

Energy minimization calculations were performed using the
conjugate gradient method for system sizes ranging from 64 to
64,000 atoms. The smallest of these systems is comparable to
moderate-sized \emph{ab initio} calculations. (We are unaware of
point defect calculations with \emph{ab initio} methods having
more than 512 atoms.) The computational cells were cubic in shape.
Separate calculations were run with with periodic and free
boundary conditions respectively. In each case the perfect crystal
was equilibrated from randomized initial conditions for the atoms.
Subsequently, the central atom was removed to model a vacancy and
the system was allowed to equilibrate again. The simulations in
periodic cells were run at zero average normal traction over each
face by including a Lagrange multiplier that is energy-conjugate
to the cell size. At each iteration of the algorithm the total
energy was minimized and the cell size was varied to obtain zero
average normal traction on each of the six faces.

The energy minimization calculations yielded a mean formation
volume over the different system sizes of
$\mathrm{tr}[\bV^\mathrm{f}] = -13.8$ \AA$^3$ in the minimum
energy configuration. This is to be compared with the atomic
volume of silicon which is $\Omega = 20$ \AA$^3$, implying a
relaxation volume of $\mathrm{tr}[\bV^\mathrm{r}] = -33.8$ \AA$^3$
according to (\ref{barnett5}).

\noindent\textbf{Remark 2}: The value of
$\mathrm{tr}[\bV^\mathrm{f}] = -13.8$ \AA$^3$ suggests a rather
strong relaxation in silicon, but is typical for the
Stillinger-Weber model, which has previously been reported to
result in large relaxations of the vacancy's nearest-neighbor
atoms \citep{balamaneetal92}. While the quantitative result is not
expected to be physically accurate, the methodology established in
this paper is of greater importance.

\subsubsection{Tensorial form of $\bV^\mathrm{r}$}\label{vrtensor}

The atomistic calculations also yield the tensorial form of
$\bV^\mathrm{r}$. The definition
\begin{displaymath}
V^\mathrm{r}_{ij} = \int\limits_{B_\mathrm{crys}}\varepsilon_{ij}
\mathrm{d}V
\end{displaymath}

\noindent can be rewritten as
\begin{displaymath}
V^\mathrm{r}_{ij} =
\int\limits_{B_\mathrm{crys}}\frac{1}{2}\left(u_{i,j} +
u_{j,i}\right) \mathrm{d}V
\end{displaymath}

\noindent where $\bu$ now is the displacement field of relaxation
around the defect obtained from the atomistic calculation. Gauss'
Theorem then gives
\begin{equation}
V^\mathrm{r}_{ij} =
\int\limits_{S_\mathrm{crys}}\frac{1}{2}\left(u_i n_j + u_j
n_i\right) \mathrm{d}A.\label{vrtensoreq}
\end{equation}

With either periodic or traction-free boundary conditions, since
the normal, $\bn$, is fixed for a given surface of the
computational cell, it follows that the integral in
(\ref{vrtensoreq}) reduces to the symmetric dyadic product of the
areal average of each displacement component, $\int\bu
\mathrm{d}A$, and $\bn$. The symmetric distribution of the
components $u_j$ on a face with non-zero normal component $n_i$,
for $j \neq i$ ensures that $\bV^\mathrm{r}$ is a diagonal tensor.
Symmetry with respect to the three basis vectors,
$\{\be_1,\be_2,\be_3\}$, then ensures isotropy of
$\bV^\mathrm{r}$. These numerical results were borne out in the
atomistic calculations to a reasonable degree of accuracy. The
tensors $\bV^\mathrm{r}$, obtained for 512 and 64,000 atom
calculations using this formula, are recorded in Appendix
\ref{append}. The averages and standard deviations over a number
of samples are reported.

Equation (\ref{vrtensoreq})  can be applied to any atomistic
calculation to determine $V^\mathrm{r}_{ij}$. In combination with
(\ref{barnett5}) it uniquely determines the defect dipole tensor,
$D_{ij}$. If the atomistic calculation does not involve
traction-free boundary conditions, the discrete analogue of the
surface integral in (\ref{barnett5}) can be determined very easily
from the atomistic results by an obvious generalization of the
procedure used below in Section \ref{boundefsymm}. This
establishes the thermodynamically-correct relation for
$V^\mathrm{f}_{ij}$ or $V^\mathrm{r}_{ij}$. The defect dipole
tensor can then be calculated in closed form using
(\ref{barnett5}) and (\ref{vrtensoreq}).

\subsection{Boundary conditions and defect
symmetry in determining
$\mathrm{tr}[\bV^\mathrm{r}]$}\label{boundefsymm}

In this section we evaluate the influence of the integral in
(\ref{barnett5}) on the value obtained for the scalar relaxation
volume, $\mathrm{tr}[\bV^\mathrm{r}]$. We consider the cubic
computational cells used in our atomistic calculations. The cell
faces were aligned with the cubic crystal directions. We emphasize
that the results of this subsection hold where the theory of
linear elasticity remains valid. One may therefore expect that
they will hold in the far-field of the point defect.

The contribution of the surface integral in (\ref{barnett5}) needs
to be evaluated on any one face only since the contributions from
the remaining faces are equal by symmetry. Consider the
$[1\,0\,0]$ face, denoted by $S_1$ with $\bx = \langle l, x_2,
x_3\rangle^\mathrm{T}$, where $l$ is the half-length of the
computational cell and $-l \le x_2, x_3 \le l$. The unit outward
normal to this face is $\bn_1 = \langle 1, 0,
0\rangle^\mathrm{T}$. The components of the traction vector
$\bt^\mathrm{P} = \Bsigma^\mathrm{P}\bn_1$, for zero average
normal stress, satisfy the following relations for an isotropic
defect:
\begin{equation}
\int\limits_{S_1}t^\mathrm{P}_1 \mathrm{d}A = 0,\quad
t^\mathrm{P}_2 = 0,\;t^\mathrm{P}_3 = 0\;\mbox{on}\; S_1,
\label{traction}
\end{equation}

\noindent The integral condition on $t^\mathrm{P}_1$ follows from
the zero average normal stress condition. (Hereafter, the phrase
``periodic boundary conditions'' will imply the zero average
normal traction condition also, unless explicitly stated
otherwise.) The pointwise conditions on the shear components
$t^\mathrm{P}_2$ and $t^\mathrm{P}_3$ follow since the shear
stress components $\sigma_{21}$ and $\sigma_{31}$ are zero on
$S_1$, which in turn follows from symmetry across the periodic
boundary and isotropy of the defect. Therefore, the cubic
anisotropy of silicon makes the trace of the boundary integral in
(\ref{barnett5}) vanish when evaluated on $S_1$:
\begin{equation}
\frac{1}{\mathbb{C}_{1111} + 2\mathbb{C}_{1122}}\int\limits_{x_2 =
-l}^l\int\limits_{x_3=-l}^l \left(t^\mathrm{P}_1 l +
t^\mathrm{P}_2 x_2 + t^\mathrm{P}_3 x_3\right) \mathrm{d}A = 0
\label{boundaryint}
\end{equation}

\noindent by Equation (\ref{traction}). Therefore, according to
linear elasticity, the periodic boundary conditions result in the
same scalar relaxation volume as the traction-free boundary,
provided the defect is isotropic and located at the center of the
periodic cell. From (\ref{barnett5}) we have
\begin{displaymath}
\mathrm{tr}[\bV^\mathrm{f}] = -\frac{1}{\mathbb{C}_{1111} +
2\mathbb{C}_{1122}}\mathrm{tr}[\bD] + \Omega,
\end{displaymath}

\noindent for silicon with cubic anisotropy and an isotropic point
defect. Therefore, according to linear elasticity, the calculation
of $\mathrm{tr}[\bV^\mathrm{f}]$ is also unaffected by the choice
of periodic or traction-free boundary conditions.

With $\mathrm{tr}[\bV^\mathrm{r}] = -33.8$ \AA$^3$ from the
atomistic calculation, $\mathbb{C}_{1111} = 1.616\times 10^{11}$
Pa and $\mathbb{C}_{1122} = 0.816\times 10^{11}$ Pa for
Stillinger-Weber silicon, \citep{balamaneetal92} this gives a
dipole strength $\mathrm{tr}[\bD] = 36.594\times 10^{-19}$ N-m for
the isotropic vacancy in the Stillinger-Weber model.

\begin{figure}[ht]
\centering
\psfrag{V}{\small$\mathsf{tr}[\mathsf{\boldmath{V}}^\mathsf{f}]$
\textsf{\AA}$^\mathsf{3}$} \psfrag{N}{\small\textsf{No. of atoms}}
\includegraphics[width=10cm]{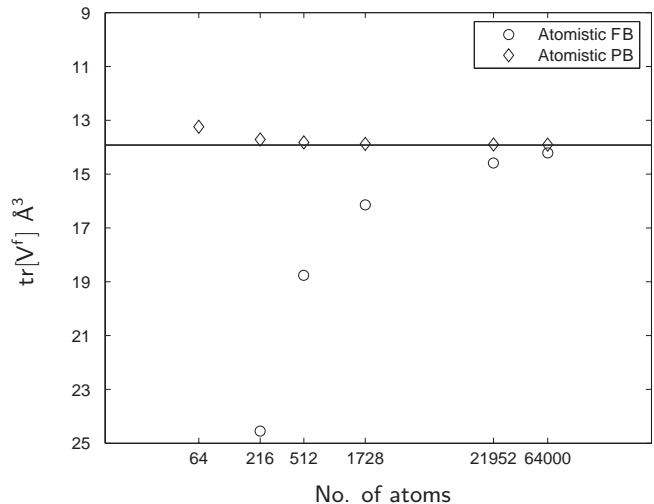}
\vspace{-1.75cm} \caption{Comparison of formation volume
calculated by atomic relaxation for traction-free boundary
conditions (Atomistic FB) and periodic boundary conditions
(Atomistic PB).} \label{vfatomicdatafig}
\end{figure}

Figure \ref{vfatomicdatafig} presents atomistic data for
$\mathrm{tr}[\bV^\mathrm{f}] = \mathrm{tr}[\bV^\mathrm{r}] +
\Omega$ as a function of number of atoms used in the calculations.
Several calculations were performed for each choice of system size
(number of atoms). These simulations were started from randomized
initial positions of atoms. Each data point represents the
formation volume of the calculation with the lowest energy for the
given number of atoms. The relative independence of
$\mathrm{tr}[\bV^\mathrm{f}]$ obtained by using periodic boundary
conditions bears out the result in (\ref{boundaryint}). The
results from atomistic simulations with traction-free boundaries
are also shown. The relatively slow convergence of these results
with system size arises from the fact that the atoms on the
surface are not identical in bonding coordination to the bulk
atoms. This also leads to the departure from the elasticity result
that $\mathrm{tr}[\bV^\mathrm{f}]$ is independent of the nature of
the boundary condition (periodic/free). This discrepancy becomes
negligible for systems having 64,000 atoms.\footnote{However, for
systems much bigger than 64,000 atoms, the sheer number of atoms
introduces many more spurious minima in the energy minimization
calculations. The conjugate gradient algorithm tends to get
``trapped'' in these spurious minima, leading to larger errors in
the relaxation volumes that are obtained.\label{spuriousmin}} The
implication is that independence from type of boundary condition
is obtained only in the larger systems where elastic effects
dominate.

\noindent \textbf{Remark 3}: The result of an atomistic
calculation for $\mathrm{tr}[\bV^\mathrm{f}]$ has been shown to be
independent of the shape and location of the boundary relative to
isotropic point defects centered in a computational cell with
periodic boundary conditions. It is therefore independent of the
size of the crystal and depends only on $\bD$, the elasticity of
the crystal, and the atomic volume. This type of boundary
condition is therefore well-suited for volume change calculations
of isotropic defects.

\section{The strain energy of formation} \label{uform}

The observed dependence of $\bV^\mathrm{f}$ on boundary conditions
in (\ref{barnett5}) motivates a similar examination of the
dependence of the strain energy of formation. In the continuum
setting it proves convenient to begin such an analysis with an
infinite crystal and subsequently subject it to cutting and
welding operations in the manner of \citet{eshelby57,eshelby61} in
order to address finite crystals. Let $B_\mathrm{crys}$ now denote
a simply-connected finite subset of the infinite crystal enclosing
the center of contraction or expansion. Let $\bn$ denote the unit
outward normal to $B_\mathrm{crys}$. Let the strain energy in the
region $B_\mathrm{crys}$ be $\mathscr{U}^\mathrm{f}_{\infty_V}$,
and the strain energy of the infinite crystal be
$\mathscr{U}^\mathrm{f}_\infty$. These energies are related as
follows:
\begin{displaymath}
\mathscr{U}^\mathrm{f}_\infty = \mathscr{U}^\mathrm{f}_{\infty_V}
+ \int\limits_{\mathbb{R}^3\backslash
B_\mathrm{crys}}\frac{1}{2}\sigma^\infty_{ij}\varepsilon^\infty_{ij}\mathrm{d}V.
\label{engyinftyV}
\end{displaymath}

\noindent Applying the Divergence Theorem to the integral, and
noting that $\sigma^\infty_{ij,j} = 0$ exterior to
$B_\mathrm{crys}$, leads to
\begin{equation}
\mathscr{U}^\mathrm{f}_\infty = \mathscr{U}^\mathrm{f}_{\infty_V}
-\int\limits_{S_\mathrm{crys}}\frac{1}{2}\sigma^\infty_{ij} n_j
u^\infty_i\mathrm{d}V. \label{engyinf}
\end{equation}

We will not pursue an explicit expression for
$\mathscr{U}^\mathrm{f}_\infty$ or
$\mathscr{U}^\mathrm{f}_{\infty_V}$. Instead we will relate them
to the strain energy of a finite crystal with boundary
$S_\mathrm{crys}$.

\subsection{Finite crystal with traction-free boundaries}\label{uformfbc}

Now consider turning $B_\mathrm{crys}$ into a finite crystal with
traction-free boundaries. This can be achieved by holding the
traction fixed at $\sigma^\infty_{ij}n_j$ on $S_\mathrm{crys}$,
cutting along this surface, and then applying a traction
$\sigma^\mathrm{F}_{ij}n_j = -\sigma^\infty_{ij}n_j$ to
$S_\mathrm{crys}$, where $\sigma^\mathrm{F}_{ij}$ is the image
stress field so produced in $B_\mathrm{crys}$. Let the
corresponding displacement field be denoted by $u^\mathrm{F}_i$,
and its strain field by $\varepsilon^\mathrm{F}_{ij} =
\mathbb{C}^{-1}_{ijkl}\sigma^\mathrm{F}_{kl}$. The relation
between the strain energy in this configuration,
$\mathscr{U}^\mathrm{f}_{\mathrm{F}_V}$, and
$\mathscr{U}^\mathrm{f}_{\infty_V}$ is
\begin{displaymath}
\mathscr{U}^\mathrm{f}_{\infty_V} =
\mathscr{U}^\mathrm{f}_{\mathrm{F}_V} +
\int\limits_{B_\mathrm{crys}}\frac{1}{2}\left(-\sigma^\mathrm{F}_{ij}\right)\left(-\varepsilon^\mathrm{F}_{ij}\right)
\mathrm{d}V,
\end{displaymath}

\noindent where we have used the fact that the stress field
$-\sigma^\mathrm{F}_{ij}$ applied to the traction-free crystal
results in a strain field $-\varepsilon^\mathrm{F}_{ij}$ in
$B_\mathrm{crys}$, returning it to the configuration with energy
$\mathscr{U}^\mathrm{f}_{\infty_V}$. Furthermore, we have used the
result that there is no interaction energy between an external
traction, $-\sigma^\mathrm{F}_{ij} n_j$, and the internal stress,
$\sigma^\mathrm{F}_{ij} + \sigma^\infty_{ij}$, since the internal
stress state is traction free [$(\sigma^\mathrm{F}_{ij} +
\sigma^\infty_{ij})n_j = 0$ by construction] at the boundary,
$S_\mathrm{crys}$.

\noindent Using integration by parts, the Divergence Theorem, and
the fact that $\sigma^\mathrm{F}_{ij,j} = 0$ this is reduced to
\begin{equation}
\mathscr{U}^\mathrm{f}_{\infty_V} =
\mathscr{U}^\mathrm{f}_{\mathrm{F}_V}
 + \int\limits_{S_\mathrm{crys}}\frac{1}{2}\sigma^\mathrm{F}_{ij}n_j
u^\mathrm{F}_i \mathrm{d}A.\label{engyfbc0}
\end{equation}

From (\ref{engyinf}) and (\ref{engyfbc0}) combined with the result
$\sigma^\mathrm{F}_{ij} n_j = -\sigma^\infty_{ij}n_j$ at
$S_\mathrm{crys}$, we have
\begin{equation}
\mathscr{U}^\mathrm{f}_\infty =
\mathscr{U}^\mathrm{f}_{\mathrm{F}_V} -
\int\limits_{S_\mathrm{crys}}\frac{1}{2}\sigma^\infty_{ij}n_j
\left(u^\mathrm{F}_i +
u^\infty_i\right)\mathrm{d}A.\label{engyfbc}
\end{equation}

\subsection{Finite crystal of cubic shape and periodic
boundaries}\label{uformpbc}

Consider a cube-shaped domain, $B_\mathrm{crys}$. Periodic
boundary conditions imposed on the displacement at constant cell
volume imply that the total displacement vanishes on
$S_\mathrm{crys}$. This state is obtained by subjecting the final
configuration with traction-free boundaries to the additional
boundary displacement $u^\mathrm{P}_i = -u^\infty_i -
u^\mathrm{F}_i$ at $S_\mathrm{crys}$. Let
$\varepsilon^\mathrm{P}_{ij}$ be the corresponding strain field
over $B_\mathrm{crys}$, and $\sigma^\mathrm{P}_{ij} =
\mathbb{C}_{ijkl}\varepsilon^\mathrm{P}_{kl}$ be the stress field
in $B_\mathrm{crys}$. Now, the strain energy of the crystal is
\begin{displaymath}
\mathscr{U}^\mathrm{f}_{\mathrm{P}_V} =
\mathscr{U}^\mathrm{f}_{\mathrm{F}_V} +
\int\limits_{B_\mathrm{crys}}
\frac{1}{2}\sigma^\mathrm{P}_{ij}\varepsilon^P_{ij} \mathrm{d}V,
\end{displaymath}

\noindent where the absence of interaction energy between
$\sigma^\mathrm{P}_{ij}$ and the internal stress with
$(\sigma^\mathrm{F}_{ij} + \sigma^\infty_{ij})n_j = 0$ on
$S_\mathrm{crys}$ has been used. Using integration by parts, the
Divergence Theorem, and the fact that $\sigma^\mathrm{P}_{ij,j} =
0$ we have
\begin{equation}
\mathscr{U}^\mathrm{f}_{\mathrm{P}_V} =
\mathscr{U}^\mathrm{f}_{\mathrm{F}_V} +
\int\limits_{S_\mathrm{crys}} \frac{1}{2}\sigma^\mathrm{P}_{ij}n_j
u^P_{i} \mathrm{d}A.\label{engypbc}
\end{equation}

The implications of (\ref{engyinf}--\ref{engypbc}) for the
thermodynamics of point defect formation are elaborated in
Sections \ref{sect3.2.2} and \ref{sect4} below.

\noindent\textbf{Remark 4}: We have used the symbol
$\mathscr{U}^\mathrm{f}$ for the strain energies of formation,
since, for the purely mechanical processes considered in this
paper, the change in internal energy, $\mathscr{U}^\mathrm{f}$, is
the strain energy when we only consider the crystal away from the
defect. An additional contribution to the formation energy arises
from the change in bonding at the defect, which is, of course, not
represented in elasticity.

\subsection{Numerical evaluation of strain energies of formation} \label{sect3.2.2}

The strain energy of formation with traction-free boundaries on a
cubic computational cell was numerically-evaluated as follows: The
stress at the boundaries of the cubic computational cell embedded
in an infinite crystal was obtained from (\ref{strain}) and
$\sigma^\infty_{ij} = \mathbb{C}_{ijkl}\varepsilon^\infty_{kl}$.
The resulting force distribution on a surface, $S_\alpha$, of the
computational cell was considered, where the surface normal was
denoted by $\bn_\alpha
=\frac{\alpha}{\vert\alpha\vert}\be_{\vert\alpha\vert},\; \alpha =
\{\pm1,\pm2,\pm3\}$. The subscript $(\bullet)_{\vert\alpha\vert}$
denotes $(\bullet)_1$, $(\bullet)_2$ or $(\bullet)_3$. The force
distribution was represented by point forces $T_{i_M}(\bx_M) =
\sigma^\infty_{ij}(\bx_M)n_{\alpha_j} A_M$, with $A_M,\; M =
1,\dots N_\mathrm{nd}^2$ being the area associated with
$N_\mathrm{nd}^2$ points, $\bx_M$, on each surface of the
computational cell. These points subsequently defined the surface
nodes of a finite element discretization of the computational cell
with $N_\mathrm{nd}^3$ nodes. The finite element mesh had
equal-sized, trilinear, cubic elements. For such a mesh, $A_M$
equals the area of each face of the cubic elements for a point
that does not lie on an edge or vertex of the computational cell;
$A_M$ equals half the area of each face of the cubic elements for
a point on an edge; and $A_M$ equals quarter the area of each face
of the cubic elements for a point on a vertex.

Image forces were applied on the nodes of $S_\alpha$ by reversing
the above-computed nodal forces so that, when superposed on the
tractions $\sigma^\infty_{ij}n_{\alpha_j}$, the resulting field at
the boundary nodes was traction-free. The displacement field
$u^\mathrm{F}_i$ due to these nodal image forces on the boundary
was obtained as the finite element displacement solution, and the
integral in (\ref{engyfbc}) was calculated as
\begin{equation}
\mathscr{U}^\mathrm{f}_{\infty} -
\mathscr{U}^\mathrm{f}_{\mathrm{F}_V} =
-\int\limits_{S_\mathrm{crys}}\frac{1}{2}\sigma^\infty_{ij}n_j
\left(u^\mathrm{F}_i + u^\infty_i\right) \mathrm{d}A =
-\frac{1}{2}\sum\limits_{\underset{\alpha\neq 0}{\alpha=-3}
}^3\sum\limits_{n=1}^{N_\mathrm{nd}^2}T_{i_n}\left(
u^\mathrm{F}_{i_n} + u^\infty_{i_n}\right).\label{approxfbcint}
\end{equation}

\noindent This is the difference in energies between the infinite
crystal and the traction-free boundary case. The calculations were
carried out for $\mathrm{tr}[\bD] = 36.594\times 10^{-19}$ N-m.

The difference between strain energies of the periodic boundary
and free boundary cases, $\mathscr{U}^\mathrm{f}_{\mathrm{P}_V} -
\mathscr{U}^\mathrm{f}_{\mathrm{F}_V}$ is given by the integral in
(\ref{engypbc}). Recall that in the atomistic calculations with
periodic boundary conditions, the computational cell was also
relaxed to have zero average normal traction, and pointwise zero
shear tractions on the surfaces. In the finite element
calculations this was achieved by applying the following traction
field over the surface $S_\alpha$:
\begin{equation}
\overline{t}^\mathrm{R}_i =\left\{\begin{array}{c@{\quad,\quad}l}
    \frac{1}{\mathrm{Area}(S_\alpha)}\int\limits_{S_\alpha}
\left(-\sigma^\mathrm{P}_{ij}n_{\alpha_j}\right)
\mathrm{d}A\, & \mbox{if}\;i=\vert\alpha\vert, \\
    -\sigma^\mathrm{P}_{ij}n_{\alpha_j}  &\mbox{if}\; i\neq\vert\alpha\vert.
  \end{array}\right.
\label{pbctrac}
\end{equation}

\noindent Note that while $\overline{t}^\mathrm{R}_i$ is constant
over the surface $S_\alpha$ for $i = \vert\alpha\vert$, it varies
over $S_\alpha$ for $i\neq \vert\alpha\vert $.

Let the displacement field $u^\mathrm{R}_i$ result from
application of traction $\overline{t}^\mathrm{R}_i$ defined as in
(\ref{pbctrac}) over each surface, $S_\alpha$. Since
$\overline{t}^\mathrm{R}_i$ is constant over $S_\alpha$ for $i =
\vert\alpha\vert$, and the shear components do not give rise to
normal strains in a cubic material, it follows that
$u^\mathrm{R}_i$ is constant over $S_\alpha$ for $i =
\vert\alpha\vert$.

In order to extend (\ref{engypbc}) to the periodic boundary
condition calculation with zero average normal traction, the
integral in that equation was replaced by
\begin{displaymath}
\sum\limits_{\underset{\alpha\neq
0}{\alpha=-3}}^{3}\int\limits_{S_\alpha}
\left(\overline{t}^\mathrm{R}_{\alpha_i} +
\sigma^\mathrm{P}_{ij}n_{\alpha_j}\right)\left(-u^\infty_i -
u^\mathrm{F}_i + u^\mathrm{R}_i\right)\mathrm{d}A,
\end{displaymath}

\noindent where $-u^\infty_i -u^\mathrm{F}_i = u^\mathrm{P}_i$.
From (\ref{pbctrac}) and $u^\mathrm{R}_i = \mathrm{constt}$ on
$S_\alpha$ for $i = \vert\alpha\vert$, this integral is
\begin{equation}
\mathscr{U}^\mathrm{f}_{\mathrm{P}_V} -
\mathscr{U}^\mathrm{f}_{\mathrm{F}_V} =
\sum\limits_{\underset{\alpha\neq
0}{\alpha=-3}}^{3}\int\limits_{S_\alpha}
\left(\overline{t}^\mathrm{R}_i + \sigma^\mathrm{P}_{ij}
n_{\alpha_j}\right)\left(-u^\infty_i -
u^\mathrm{F}_i\right)\mathrm{d}A. \label{engypbc1}
\end{equation}

The finite element displacement solution obtained for
$-u^\mathrm{F}_i$ was added to the infinite crystal displacement
field $-u^\infty_i$. This field,$- u^\infty_i - u^\mathrm{F}_i$,
evaluated on the boundaries, $S_\alpha$, was then reapplied as a
displacement boundary condition to the finite element mesh to
obtain the surface traction field
$\sigma^\mathrm{P}_{ij}n_{\alpha_j}$. Then (\ref{pbctrac}) and
(\ref{engypbc1}) gave the required energy difference.

\begin{figure}[ht]
\centering \psfrag{E}{\small$\mathscr{U}^\mathsf{f}_\infty -
\mathscr{U}^\mathsf{f}_{(\bullet)}$ \textsf{(eV)}}
\psfrag{N}{\small\textsf{No. of atoms}}
\includegraphics[width=10cm]{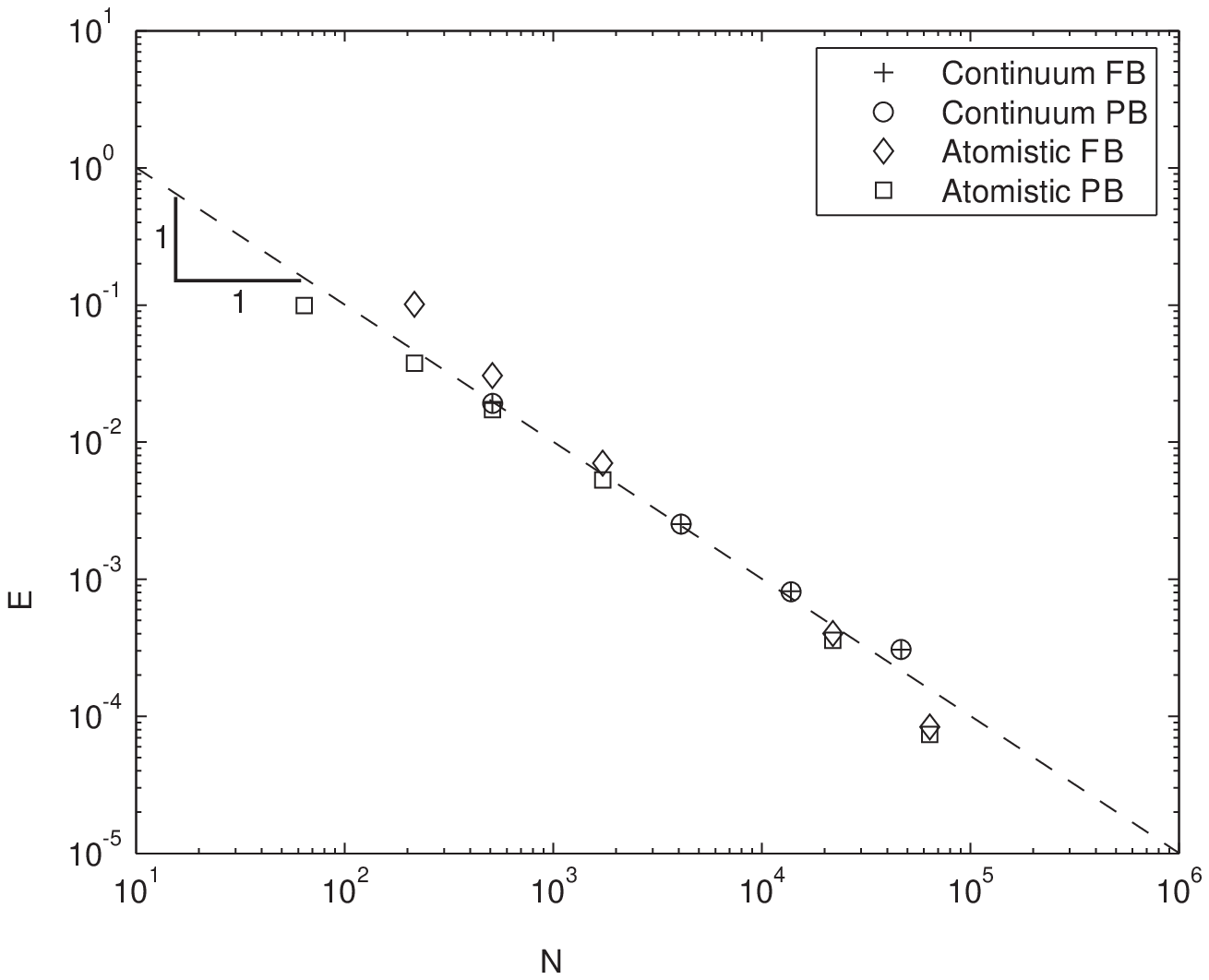}
\caption{Comparison of strain energies of formation calculated by
atomic relaxation and elasticity for periodic (PB) and
traction-free (FB) boundary conditions.} \label{energyplots}
\end{figure}

Figure \ref{energyplots} is a comparison of the energy differences
$\mathscr{U}^\mathrm{f}_\infty -
\mathscr{U}^\mathrm{f}_{\mathrm{F}_V}$ and
$\mathscr{U}^\mathrm{f}_\infty -
\mathscr{U}^\mathrm{f}_{\mathrm{P}_V}$ obtained with the atomistic
and elasticity calculations. The actual value of
$\mathscr{U}^\mathrm{f}_\infty$, has not been obtained directly
from elasticity. However, the integrals in (\ref{approxfbcint})
and (\ref{engypbc1}) both are $\mathcal{O}(r^{-3})$, where $2r$ is
the characteristic linear dimension of the computational cell.
Therefore, $\mathscr{U}^\mathrm{f}_{\mathrm{F}_V}$ and
$\mathscr{U}^\mathrm{f}_{\mathrm{P}_V}$ converge to
$\mathscr{U}^\mathrm{f}_\infty$ as $r \to \infty$. Since
$r^{-3}\propto N^{-1}$, where $N$ is the number of atoms in the
system, this convergence is confirmed in Figure \ref{energyplots}.
Analysis of (\ref{engyfbc}) and (\ref{engypbc}), or of
(\ref{approxfbcint}) and (\ref{engypbc1}) indicates that
$\mathscr{U}^\mathrm{f}_{\mathrm{F}_V}$ and
$\mathscr{U}^\mathrm{f}_{\mathrm{F}_V}$ converge to
$\mathscr{U}^\mathrm{f}_\infty$ from below. Figure
\ref{energyplots} confirms this prediction, since the energy
differences $\mathscr{U}^\mathrm{f}_\infty -
\mathscr{U}^\mathrm{f}_{(\bullet)}$ are all positive. The trends
are replicated by the atomistic calculations, in which
$\mathscr{U}^\mathrm{f}_{\mathrm{F}_V}$ and
$\mathscr{U}^\mathrm{f}_{\mathrm{P}_V}$ are directly calculated.
The asymptotic value to which
$\mathscr{U}^\mathrm{f}_{\mathrm{F}_V}$ and
$\mathscr{U}^\mathrm{f}_{\mathrm{P}_V}$ converge in the atomistic
calculations has been used as $\mathscr{U}^\mathrm{f}_\infty$.

The discrepancy between linear elasticity and atomistic
calculations for the smaller cells (e.g. 512 atoms) in Figure
\ref{energyplots} is most probably due to the highly nonlinear
deformations around the vacancy that dominate at these small
scales. The 512-atom computational cell is made up of $4\times
4\times 4$ Si unit cells. The boundary is only 10.86 \AA\ away
from the vacancy. At such small scales linear elasticity provides
a poor estimate of the strain energy of the highly distorted
lattice around the vacancy.

The difference between the energies of the traction-free and
periodic boundaries with the atomistic calculations in Figure
\ref{energyplots} most probably is a manifestation of an
interaction energy in the near field of the defect: The periodic
boundary configuration effectively applies a boundary traction to
the traction-free configuration, as explained above. This external
stress field interacts with the internal stress of the
traction-free boundary configuration. Such an interaction energy,
however, vanishes according to the theory of linear elasticity.
This result has been employed in Sections \ref{uformfbc} and
\ref{uformpbc}, and is possibly responsible for the very small
differences observed between
$\mathscr{U}^\mathrm{f}_{\mathrm{F}_V}$ and
$\mathscr{U}^\mathrm{f}_{\mathrm{P}_V}$ using linear elasticity.
For the smaller cells, the strong nonlinear effect possibly
invalidates this result of linear elasticity causing the
discrepancy in $\mathscr{U}^\mathrm{f}_{\mathrm{P}_V} -
\mathscr{U}^\mathrm{f}_{\mathrm{F}_V}$ between the atomistic and
linear elasticity calculations. The trends, however, are the same:
The free and periodic boundary cases converge from below to the
infinite crystal result. The periodic boundary case is higher in
strain energy due to the imposed constraint that preserves the
flat shape of the cubic cell's boundaries. This difference, while
small, exists in the elasticity calculations also as seen in Table
\ref{tab1}.

\begin{table}[ht]
\centering \caption{Comparison of strain energy of vacancy
formation in a cubic cell with periodic and traction-free
boundaries---elasticity results. For comparison, the atomistic
calculations converge to $\mathscr{U}^\mathrm{f}_\infty = 2.8237$
eV. Note that the cell sizes, expressed as number of Si atoms that
make up a crystal of the same size, do not correspond exactly to
the cell sizes actually used in the atomistic
calculations.}\label{tab1}
\begin{tabular}{ll}
\hline
\multicolumn{1}{c}{Cell size in number of atoms} & \mbox{$\mathscr{U}^\mathrm{f}_{PV}-\mathscr{U}^\mathrm{f}_{FV}$ ($\mu$eV)}\\
\hline
512 & 45.791\\
4096& 5.759\\
13824& 1.714\\
46656& 0.510\\
110592& 0.215\\
\hline
\end{tabular}
\end{table}

\section{Conclusions}
\label{sect4}

The central conclusions from this work are the following:

\begin{enumerate}

\item Equation (\ref{barnett5}) implies that the formation volume
of a point defect is a well-defined quantity. It is to be measured
on a crystal whose surfaces are traction-free. Defined in this
manner, it depends only on the strength of the defect, the
elasticity of the material and the atomic volume. It is
independent of the crystal's shape and size, and of the location
of the defect within the crystal.

\item On the basis of Item 1 above, Equation (\ref{barnett5}) provides an
exact quantitative measure of the appropriateness of any class of
boundary conditions used on an atomistic calculation to extract
formation volumes of point defects.

\item The derivation ending in Equations (\ref{garikipati1}) and
(\ref{garikipati2}) demonstrates that Eshelby's calculation
(\ref{eshelby1}) on the work done by external stress on a
transforming inclusion has an analogue for point defects. This is
an exact result. It provides a rigorous thermodynamic basis for
the concepts of formation and relaxation volume tensors that goes
beyond formal definitions of the type $\bV^\mathrm{f} =
-\partial\mathscr{G}^\mathrm{f}/\partial \Bsigma^\mathrm{ex}$.

\item The combination of (\ref{vrtensoreq}) and (\ref{barnett5}) provides a
closed-form expression for the defect dipole tensor that can be
determined from any atomistic calculation.

\item Equations (\ref{engyinf}--\ref{engypbc}) quantify the
influence of traction-free and periodic boundary conditions on the
strain energy of formation of point defects according to continuum
linear elasticity. In each case, the corresponding boundary
integrals can be calculated. Significant discrepancies are
observed between the continuum elastic and atomistic results for
the smallest system sizes (512 atoms) because the highly nonlinear
lattice distortion around the defect permeates the entire
computational cell when the latter is composed of a small number
of atoms. Linear elasticity is not an accurate theory for such
nonlinear distortions. Additionally, for atomistic systems of this
size, there appears to be an interaction energy between an
externally-applied boundary traction and an internal stress field
with vanishing traction at the boundary. According to linear
elasticity this interaction energy vanishes exactly. This could
lead to the much smaller difference in strain energies of
formation between the traction-free and periodic boundary cases
obtained via linear elasticity. Indeed, this suggests that
(\ref{engyfbc0}--\ref{engypbc}) are not accurate for systems
smaller than roughly 4000 atoms.

\item The difference between strain energies of vacancy formation
obtained with traction-free boundaries and periodic boundaries
(that are traction free on average) vanishes with increasing cell
size. Linear elasticity predicts that the scalar vacancy formation
volumes, however, are identical for an isotropic defect that is
centered in the computational cell. Yet, the atomistic results
show significant differences for the scalar formation volume for
the two types of boundary conditions (Figure
\ref{vfatomicdatafig}) at smaller cells. This discrepancy can be
attributed to the larger influence of boundary atoms' coordination
numbers for small cells, and it diminishes for the larger
calculations. We expect that this discrepancy will be enhanced,
due to elasticity effects, for the much larger class of
anisotropic point defects: vacancies with the Jahn-Teller
distortion and most interstitial configurations, as well as for
defects not located at the center of the periodic cell.

\end{enumerate}

\appendix
\section{Relaxation volume tensors from atomistic calculations}
\label{append}

Representative relaxation volume tensors obtained from the energy
minimization calculations on systems with varying numbers of
atoms, and with traction-free or periodic boundary conditions are
reported below. For the sake of brevity values are reported for
only the 512- and 64,000-atom systems. Averages and standard
deviations (SD) were calculated by running several calculations
with starting positions of atoms varying randomly from the
equilibrium positions in a perfect crystal. The larger standard
deviations for larger systems are associated with the spurious
minima discussed in Footnote \ref{spuriousmin}.

Traction-free boundary conditions with 512 atoms:

\begin{displaymath}
\bV^\mathrm{r}_\mathrm{av} = \left[\begin{array}{ccc}-13.676& 0.683& 0.695\\
0.683& -13.669& 0.678\\ 0.695& 0.678&
-13.674\end{array}\right]\,\mbox{\AA}^3,\quad \mbox{SD} = \left[\begin{array}{ccc}0.192& 0.030& 0.031\\
0.030& 0.172& 0.027\\ 0.031& 0.027&
0.151\end{array}\right]\,\mbox{\AA}^3.
\end{displaymath}

Periodic (traction-free on average) boundary conditions with 512
atoms:

\begin{displaymath}
\bV^\mathrm{r}_\mathrm{av} = \left[\begin{array}{ccc}-12.009& 0.706& 0.706\\
0.706& -12.009& 0.706\\ 0.706& 0.706&
-12.009\end{array}\right]\,\mbox{\AA}^3,\quad \mbox{SD} = \left[\begin{array}{ccc}0.001& 0.002& 0.002\\
0.002& 0.001& 0.001\\ 0.002& 0.001&
0.001\end{array}\right]\,\mbox{\AA}^3.
\end{displaymath}

Traction-free boundary conditions with 64,000 atoms:

\begin{displaymath}
\bV^\mathrm{r}_\mathrm{av} = \left[\begin{array}{ccc}-11.625& -0.022& -0.021\\
-0.022& -11.484& -0.029\\ -0.021& -0.029&
-11.421\end{array}\right]\,\mbox{\AA}^3,\quad \mbox{SD} = \left[\begin{array}{ccc}0.535& 0.178& 0.185\\
0.178& 0.308& 0.138\\ 0.185& 0.138&
0.339\end{array}\right]\,\mbox{\AA}^3.
\end{displaymath}

Periodic (traction-free on average) boundary conditions with
64,000 atoms:

\begin{displaymath}
\bV^\mathrm{r}_\mathrm{av} = \left[\begin{array}{ccc}-11.571& -0.147& -0.130\\
-0.147& -11.217& -0.054\\ -0.130& -0.054&
-11.136\end{array}\right]\,\mbox{\AA}^3,\quad \mbox{SD} = \left[\begin{array}{ccc}0.653& 0.115& 0.057\\
0.115& 0.473& 0.142\\ 0.057& 0.142&
0.363\end{array}\right]\,\mbox{\AA}^3.
\end{displaymath}

\bibliographystyle{elsart-harv}
\bibliography{mybib}

\end{document}